\documentstyle[aps,epsf]{revtex}
\newcommand{\be}{\begin{equation}}
\newcommand{\ee}{\end{equation}}
\newcommand{\beq}{\begin{eqnarray}}
\newcommand{\eeq}{\end{eqnarray}}

\begin{document}

\title{The Gluon Propagator
without lattice Gribov copies}

\author{C. Alexandrou$^{1,\>2}$, Ph. de Forcrand$^{3,\>4}$, E. Follana$^1$\\
$^1$ Department of Physics,
University of Cyprus, CY-1678 Nicosia, Cyprus\\
$^2$ Paul Scherrer Institute, CH-5232 Villigen, Switzerland \\
$^3$Inst. f\"ur Theoretische Physik, ETH H\"onggerberg, CH-8093 Z\"urich,
Switzerland\\
$^4$ CERN, Theory Division, CH-1211 Geneva 23, Switzerland}

\maketitle

\begin{abstract}
We study the gluon propagator in quenched  lattice QCD
 using the Laplacian gauge 
which is free of lattice Gribov copies. We compare
our results with those obtained in the Landau gauge on the lattice, as well
as with various approximate solutions of the Dyson Schwinger equations. 
We find a  finite value $\sim (445 \rm{MeV})^{-2}$ 
for the renormalized zero-momentum
propagator (taking our renormalization point at $1.943$ GeV),
 and a pole mass $\sim 640 \pm 140$ MeV.

\end{abstract}


\section{Introduction}
Over the last twenty years,  widely different
 conjectures have been proposed for the 
infrared behaviour of the gluon propagator.
Although it is a gauge dependent quantity,   
it can be discussed in a given gauge. Even within the same gauge,
the proposals for the infrared dependence differ drastically~\cite{Mandula}. 
We mainly summarize here the
results that are given in the literature within the Landau gauge, since
that gauge is widely  used  in studies of Dyson Schwinger equations
(DSE) as well as in lattice QCD.
Early predictions were obtained by solving
approximately the DSE. Mandelstam~\cite{Mandelstam}
obtained a solution of a set of truncated DSE equations with 
an infrared behaviour of the form $(q^2)^{-2}$ 
for the gluon propagator. Such an infrared enhancement 
was shown, if obtained in any gauge, to lead to an
area law for the Wilson loop~\cite{West} and thus to be sufficient 
for confinement. Infrared enhancement was assumed in various phenomenological
studies~\cite{BBZ} and corroborated by later studies of DSE with 
refined approximations~\cite{BP}.
 A different perspective was taken by
 Gribov~\cite{Gribov}, who showed that 
avoiding gauge copies one would obtain
a gluon propagator which vanishes in the infrared in the Landau and
Coulomb gauges, of the form 
\be
 D(q^2) \sim \frac{q^2}{q^4 + m^4} \quad.
\ee
An infrared suppressed  behaviour was advocated
 by Stingl~\cite{Stingl}, and recently by others~\cite{LHA}, as a possible
solution to DSE. Following a procedure  similar to that by Gribov,
 Zwanziger~\cite{Zwanziger} gave arguments to show that,
on the lattice, for any finite spacing in the limit
of infinite volume, $ D(q^2=0) =0$.

We will also consider in this work 
 the parametrization deduced by Cornwall~\cite{Cornwall}
using a resummation of
Feynman graphs which leads to gauge-invariant amplitudes. The gluon
propagator is obtained as a solution to this special set of DSE where
the claim is that the only gauge dependence appears in the free part. 
Cornwall's solution, in
addition to fulfilling the Ward identities, allows a dynamical mass 
generation. 
Thus this formulation has the additional attractive feature that 
the gluon mass vanishes in the ultraviolet as required perturbatively.
Since the self-energy obtained by Cornwall is claimed to be gauge-independent,
we will use his model to fit 
the propagator both in the Landau and in the Laplacian
gauge.

In contrast to all the approaches described above, lattice QCD provides
a framework for the calculation of the gluon propagator
starting directly from  the QCD Lagrangian and  can thus yield a conclusive
result. Attempts to calculate the gluon propagator  started 
more than ten years ago~\cite{MO,Gupta} on rather small lattices. These 
early results could be interpreted in terms of a massive scalar
propagator, but confirmed the expectation that a Lehmann-K\"allen
representation is not applicable: positivity of the transfer matrix is lost
after the non-local gauge fixing. Results on larger lattices
were accounted for by assuming a positive anomalous
dimension~\cite{Marenzoni}. Recently, a detailed study
of the gluon propagator on very large lattices~\cite{Leinweber} has been
performed, which makes an impressive effort towards bringing under control
errors due to the finite lattice spacing and to the finite lattice volume.
However, up to now, all  lattice studies have used a similar implementation of
the Landau gauge on the lattice. Gauge-fixing is accomplished by using a
local iterative procedure which
 identifies local stationarity, but in general
fails to determine the global extremum. Which local extremum 
(``lattice Gribov copy'') is selected depends on the starting condition. These
lattice Gribov copies cannot be eliminated. In this situation, their effect
has repeatedly been claimed to be small~\cite{Grib_effects}. As discussed in 
Section 3, we are not convinced by such claims. 
Results obtained so far will have a safer foundation if the effects of
 lattice Gribov copies
are better understood. 

In this work we address the problem of Gribov copies. We use
a different gauge condition, which produces a smooth gauge field like the
Landau gauge, but which specifies the gauge uniquely:
no ambiguity arises due the lattice gauge fixing procedure. This
is accomplished by using the Laplacian gauge~\cite{VW}. 
The motivation and  implementation 
of this gauge are given in section 3.

We calculate the gluon propagator in quenched QCD on lattices of sizes $8^4,
16^4$ and $16^3\times 32$ at $\beta=5.8$ and $6.0$, in an attempt to 
study its zero-temperature behaviour.
Our procedure can be extended straightforwardly to finite temperature where the
infrared behaviour of the propagator yields the chromo-electric and
chromo-magnetic screening masses.
The results that we obtain, within the Laplacian  gauge, show 
the same  ultraviolet behaviour
as in Landau gauge. However,
there are significant modifications in the infrared. In particular we
find that the zero-momentum propagator is finite,
 obeys scaling, and becomes volume independent 
for large enough volumes. It should not however be used as
a definition of the gluon mass, since the zero-momentum
limit of the propagator is gauge dependent. 
It is simply a measure of the susceptibility of the gauge-fixed field
$A_\mu$ in the Laplacian gauge.
A quantity which instead can be shown to be gauge independent to all orders in
perturbation theory is the pole mass of the transverse
part $D(q^2)$ of the propagator~\cite{KKR}. 
To determine this pole if it exists at all, an extrapolation to negative $q^2$
is necessary.
We compare the inverse propagator $D^{-1}(q^2)$ in the Laplacian and the Landau
gauges.  
Using a variety of extrapolation ans\"atze, in particular a fit to 
Cornwall's model~\cite{Cornwall} which describes the momentum
dependence of our results rather well, we find that data in the Laplacian
gauge give support for the existence of a pole at a mass
of $\sim 640(140)$ MeV. Data at smaller momenta are needed to consolidate this
result.
     
Section II introduces our notation; Section III motivates and describes
our choice of the Laplacian gauge; Section IV presents our results.
They are summarized in Section V.

\section{Definition of the gluon propagator}

\noindent
The gluon propagator in the continuum is given by
\be {\cal D}^{ab}_{\mu\nu}(q) = -i \int d^4x 
\langle 0|T[A^a_\mu(x) A^b_\nu(0)]|0 \rangle e^{iq.x}
\ee
This tensor can be decomposed into a transverse and a longitudinal part:
\be
{\cal D}^{ab}_{\mu\nu}(q) = \biggl(\delta{\mu\nu} -\frac{q_\mu q\nu}{q^2}\biggr)
               \>  \delta^{ab} \> D(q^2)  +
            \frac{q_\mu q_\nu}{q^2}\>\delta^{ab}\> \frac{F(q^2)}{q^2}
\ee
For a covariant gauge $F(q^2)$ reduces to a constant and corresponds to the
gauge fixing parameter $\xi$ which in the  Landau gauge  is zero.
Since we want to make a comparison with the recent
results~\cite{Leinweber}
 obtained in the Landau gauge,
we study the  transverse scalar function $D(q^2)$ which can be  extracted from
${\cal D}^{ab}_{\mu\nu}(q)$:
\be
D(q^2) = \frac{1}{3}\biggl\{\sum_\mu \frac{1}{8} 
                     \sum_a {\cal D}^{aa}_{\mu\mu}(q) \biggr\}
         -\frac{1}{3}  \frac{F(q^2)}{q^2} \quad.
\ee
$F(q^2)$ is determined by projecting the longitudinal part of 
${\cal D}^{aa}_{\mu\nu}(q)$ using the symmetric tensor $q^\mu q^\nu$.

On the lattice the dimensionless gluon field can be defined by
\be
A_{\mu}(x+\hat{\mu}/2)  =\frac{1}{2ig_0}\biggl\{\left[U_\mu(x)-U^{\dagger}_\mu(x)\right]
 -\frac{1}{3} {\rm Tr}\left[U_\mu(x)-U^{\dagger}_\mu(x)\right]\biggr\} 
 + {\cal O}(a^3)
\ee
where $a$ is the lattice spacing. 
One may consider different definitions for the gluon field $A_\mu$, accurate
to higher order in $a$. It has been found~\cite{Giusti} 
that these different definitions give rise to modifications that 
can be absorbed in the multiplicative field
renormalization constant. 

\noindent
The gluon propagator in momentum space is constructed by
taking the  discrete Fourier transform of $A_\mu$ for each colour component 
\be
A^a_\mu(q) = \sum_x e^{-i q.(x+\hat{\mu}/2)} A^a_\mu(x+\hat{\mu}/2)
\ee
where the discrete momentum $q=(q_\mu,\mu=1,..,4)$ takes values
\be
 q_\mu =\frac{2\pi}{a {\rm L}_\mu} n_\mu, \hspace*{1cm}
              n_\mu=-(\frac{1}{2}{\rm L}_\mu-1),...,(\frac{1}{2}{\rm L}_\mu)
\ee
and the momentum-space gluon propagator $D^{ab}_{\mu\nu}(q)$ is defined by
\be
{\rm V}~ \delta(q-q') D^{ab}_{\mu\nu}(q) = \langle A^a_\mu(q)A^b_\nu(-q') \rangle \quad.
\ee
with $V$ the lattice volume.  
In the ultraviolet the gluon propagator is expected to behave like $1/q^2$.
Since
on the lattice the free massless propagator behaves as
\be
D(q) = \frac{1}{\sum_\mu\left(\frac{2}{a} \>\>\sin
\frac{q_\mu a}{2}\right)^2} ,
\ee
to reduce errors due to the finite lattice spacing
we take as our momentum variable the usual
\be
\hat{q}_\mu=\frac{2}{a}{\rm sin}\frac{q_\mu a }{2} 
\ee

To relate the bare lattice propagator to the renormalized continuum propagator
$D_{\rm R}(q;\mu)$
one needs the renormalization constant $Z_3(\mu,a)$:
\be
a^2 D(qa) = Z_3(\mu,a) D_{\rm R} (q;\mu) \quad.
\label{D_R}
\ee
Imposing a renormalization condition such as
\be
D_{\rm R}(q)|_{q^2=\mu^2} = \frac{1}{\mu^2}
\label{renormalization point}
\ee
at a renormalization scale $\mu$ allows a determination of $Z_3(\mu,a)$.
Connection to other continuum renormalization schemes can then be made.

\section{Gauge fixing procedure}

\subsection{Motivation}

The gluon propagator is normally considered in Landau gauge, 
$\partial_\mu A_\mu(x) = 0 ~\forall x$. 
On the lattice, this condition becomes:
\be
F(\Omega) \equiv
\sum_{x,\mu} {\rm Re~Tr}~(\Omega(x)^\dagger U_\mu(x) \Omega(x+\hat{\mu}))
~~~~~{\rm maximum}
\label{GFF}
\ee
The gauge-fixing functional $F$ has many local maxima. To specify the
gauge uniquely, the gauge condition above refers to the {\em global}
maximum. This defines the {\em Fundamental Modular Region} (FMR) Landau gauge.
In practice however, the gauge transformation $\Omega$ is found
by an iterative local maximization of $F$, which terminates when any {\em local}
maximum has been reached.
A different gauge condition is thus implemented, which one might call the
{\em random} Landau gauge, and which depends on the details of the maximization
procedure.

It is commonly believed that the effect of choosing a local maximum of 
(\ref{GFF}) rather than the global maximum is small, so that the ``random''
Landau gauge is a good approximation to the FMR Landau gauge. 
The following argument is often presented to support this view.
 A given gauge configuration is
gauge-fixed $n$ times, each time after 
performing a random gauge transformation;
this procedure generates many gauge copies, each corresponding to the 
local maximum nearest to the random starting point along the gauge orbit.
It is observed~\cite{Cucchieri} that the difference between gluon propagators
measured on copies corresponding to the largest and the smallest values of (\ref{GFF}) 
is found to be statistically insignificant. 
A possible problem with this argument however is that
the number $n$ of gauge copies considered in such comparisons
(typically 30 or less)
is extremely small compared to the total number of local extrema of (\ref{GFF}):
for simple entropic reasons, all copies considered miss the global maximum
by similar amounts, and no reliable information can be extracted about the
gluon propagator in the global maximum configuration. 
It is therefore possible, and we believe quite likely, that the ``random''
Landau gauge and the FMR Landau gauge are significantly different.

Further evidence for this situation has recently been provided in another
gauge, the Direct Maximal Center (DMC) gauge~\cite{DMC}. Although the functional
$F_{DMC}(\Omega)$ to be maximized differs from (\ref{GFF}), a similar approach
of local iterative maximization is taken, leading to the ``random'' DMC gauge,
with similar problems. In this case
however, it is also possible to converge to a large value $\tilde{F}_L$ 
of $F_{DMC}$ by starting from a Landau gauge copy (``Landau'' DMC gauge). 
This value $\tilde{F}_L$ can then be compared with
the values obtained from $n$ random starting points. One may fit the maximum
value among $n$ copies, $\tilde{F}(n)$, by a reasonable ansatz like a 
series in $1/n$, and extrapolate to $n \rightarrow \infty$. It turns out 
that the extrapolated value falls well below $\tilde{F}_L$, which is itself
below the global maximum~\cite{Bornyakov}. 
Furthermore, the properties of the gauge-fixed field are qualitatively 
different between the ``random'' and the ``Landau'' DMC gauges: the former
confines after center projection, while the latter does not~\cite{KT}.

Since in Landau gauge as in DMC gauge, the number of local maxima is 
expected to grow
exponentially with the lattice volume, we expect a similar situation in Landau
gauge, leading to large differences between the ``random'' (local maximum) 
and the FMR (global maximum) gauges.
One might argue that this is not a problem, and 
that the local maximization of (\ref{GFF}) implements 
in the thermodynamic limit a well-defined, but {\em stochastic} gauge condition.
The relationship between that gauge condition and its perturbative version
$\partial_\mu A_\mu(x) = 0$ is unclear however.
Therefore, one should consider the possible
effects of selecting a local rather than the global maximum of (\ref{GFF})
with a great deal of caution.
This is the motivation for our study of the gluon propagator in a well-defined,
unambiguous gauge.

\subsection{$SU(3)$ Laplacian gauge fixing}

In \cite{VW}, Vink and Wiese proposed a simple method to fix the gauge
unambiguously in $SU(N)$. It uses $N$ auxiliary Higgs fields, which are chosen
as the $N$ lowest-lying eigenvectors $v^{(i)}$ of the covariant Laplacian. 
Under a local gauge transformation $\Omega(x)$, these eigenvectors transform
covariantly: $v^{(i)}(x) \rightarrow \Omega(x) v^{(i)}(x)$. Therefore, the
gauge can be fixed by requiring, at each space-time point $x$, 
$\{\Omega(x) v^{(i)}(x), i=1,..,N\}$
to have some predefined orientation in color space.
Specifically, each eigenvector $v^{(i)}(x)$ has $N$ complex color components,
so that the $N$ eigenvectors form a complex $N$ by $N$ matrix $M$.
Ref.\cite{VW} projects this matrix onto $SU(N)$ by polar decomposition:
$M = W P, W \in U(N), P = (M^\dagger M)^{1/2}$. 
The required gauge transformation is then
$\Omega(x) = e^{i \alpha} W^\dagger$, where 
$\alpha = \frac{1}{N} {\rm arg}(\det W)$.
$\Omega(x)$ rotates $M$ ``parallel'' to the identity ${\bf 1}_N$ at each 
space-time point.
The gauge is unambiguously defined, except for these gauge configurations
where some of the $N$ lowest eigenvalues are degenerate. Such configurations
are genuine Gribov copies; they never occur in practice.
This approach has been tested for $SU(2)$ and $U(1)$~
\cite{Vink} and it was shown
to reduce to the Landau gauge in the continuum limit aside from
exceptional configurations (e.g. an instanton background).
Here, we use a slightly modified procedure which requires only $(N-1)$ 
eigenvectors (2 for $SU(3)$), as follows \cite{Pepe}.

First, apply a gauge transformation $\Omega^{(1)}(x)$ which rotates $v^{(1)}(x)$
to 
$\left( \begin{array}{c}
|v^{(1)}(x)|  \\
       0      \\
       0 
\end{array} \right)$.
Five real components of the rotated $v^{(1)}(x)$ must vanish, which specifies five
constraints. Therefore $\Omega^{(1)}(x)$ is not fully specified, but has
$8-5=3$ degrees of freedom. Any satisfactory $\Omega^{(1)}$ can be used.

To completely fix the gauge, we use the second eigenvector $v^{(2)}$,
already rotated by $\Omega^{(1)}$ to 
$\left( \begin{array}{c}
v^{(2)}_1 \\
 v^{(2)}_2 \\
 v^{(2)}_3  
\end{array} \right)$.
Three additional constraints are obtained by requiring $v^{(2)}$ to be 
rotated to
$\left( \begin{array}{c}
v^{(2)}_1 \\
\sqrt{|v^{(2)}_2|^2 + |v^{(2)}_3|^2} \\
       0 
\end{array} \right)$.
This fixes the gauge completely and uniquely.

Note that the second rotation is in an $SU(2)$ subgroup, since it leaves
$v^{(2)}_1$ untouched. This indicates how
to generalize this construction to $SU(N)$: the first rotation fixes $(2N-1)$
constraints, which leaves $(N^2-1) - (2N-1) = ((N-1)^2 - 1)$ degrees of freedom,
forming a subgroup $SU(N-1)$. The next step reduces the gauge freedom to 
$SU(N-2)$, etc... down to $SU(2)$. 
It is easily seen that, in this recursive procedure, the matrix $M$ is reduced
to upper triangular form (with real positive diagonal elements) by the
rotation $\Omega(x)$. This is why the $N^{th}$ eigenvector needs not be
computed: it is only transformed by a phase, which is separately determined 
by the requirement that $\Omega(x) \in SU(N)$.
Our procedure can thus be viewed as a $QR$ decomposition of $M$.
The gauge, which is globally well-defined (provided the $N$ eigenvalues are
distinct), may be ill-defined on a sub-manifold of points $x$ where our
recursive process breaks down. It can be seen that such local gauge defects
occur at isolated points, where for $SU(3)$, 
$\sqrt{|v^{(2)}_2|^2 + |v^{(2)}_3|^2} = 0$.
The correlation of these points with instantons is studied in \cite{Pepe}.

The Laplacian gauge so defined has the great virtue of being unambiguous.
Hence it is the appropriate tool to address our concern about the effect of
local extrema of the usual Landau gauge. It also has strong similarities with
Landau gauge: it is smooth, Lorentz-symmetric, and gauge-fixes a pure gauge
lattice configuration (gauge-transformed from $U_\mu(x)={\bf 1} ~\forall x,\mu$)
back to $U={\bf 1}$. Nevertheless, it is a different gauge: its perturbative
definition is under consideration \cite{PvB}; it differs from Landau gauge most strongly
where the magnitude of the eigenvectors $|v^{(1,2)}(x)|$ becomes small.

\section{Results} 

Since most of the previous studies were performed in the Landau gauge, it is
important to compare our Laplacian-gauge propagator 
with the Landau-gauge one.
For this purpose, we have taken, for our analysis, lattice configurations 
available on the Gauge Connection database \cite{connection}, 
which had already 
been gauge fixed to Landau gauge with the usual local Over-Relaxation 
method~\cite{Landau_OR}. 
These are 200 configurations of a $16^3 \times 32$ lattice,
at $\beta=5.8$ and $6.0$ each.

\begin{figure}[t]
\epsfxsize=13truecm
\epsfysize=10truecm
\vspace*{0.3cm}
\mbox{\epsfbox{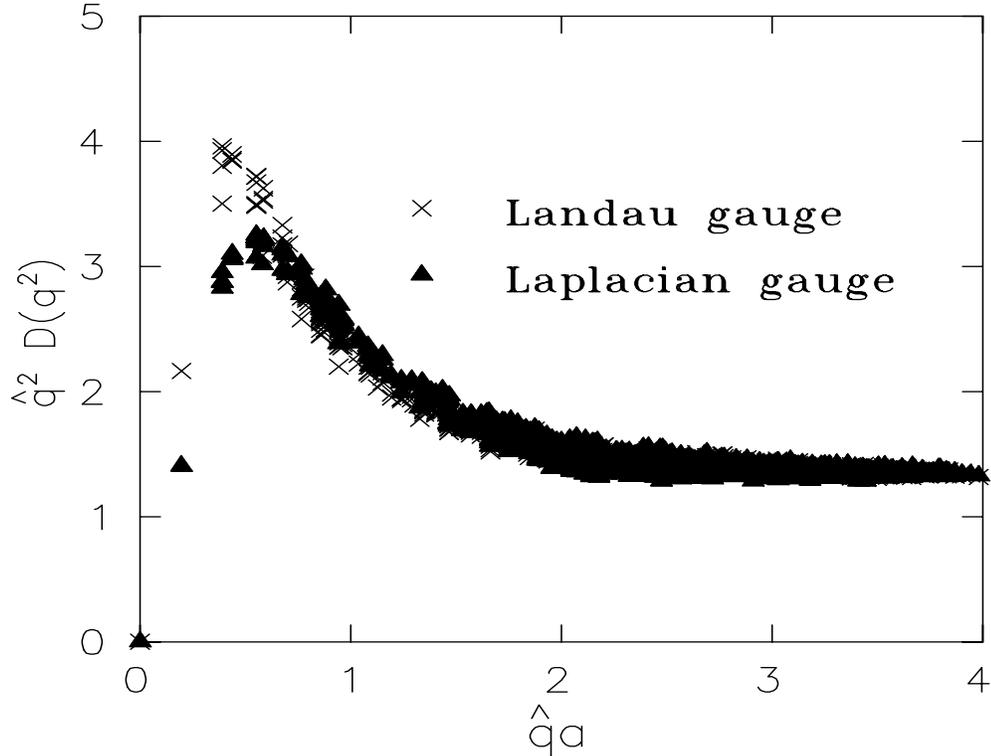}}
\caption{Comparison of the transverse 
gluon propagator times $q^2$ in Landau and Laplacian 
gauge on a $16^3\times 32$ lattice at $\beta=6.0$.}
\label{fig:compare}
\end{figure}

The transverse gluon propagator is shown in Fig.~\ref{fig:compare} 
for the two gauges at $\beta=6.0$. 
As expected the ultraviolet behaviour is identical in the two gauges,
whereas in the infrared, which is the region of interest, significant
differences are visible. Since we use a different gauge, this should not
come as a surprise.
We  show the usual quantity $\hat{q}^2 D(q^2)$.
The Laplacian propagator is clearly not as large as the Landau propagator
at low momenta.

\begin{figure}[b]
\epsfxsize=13truecm
\epsfysize=10truecm
\mbox{\epsfbox{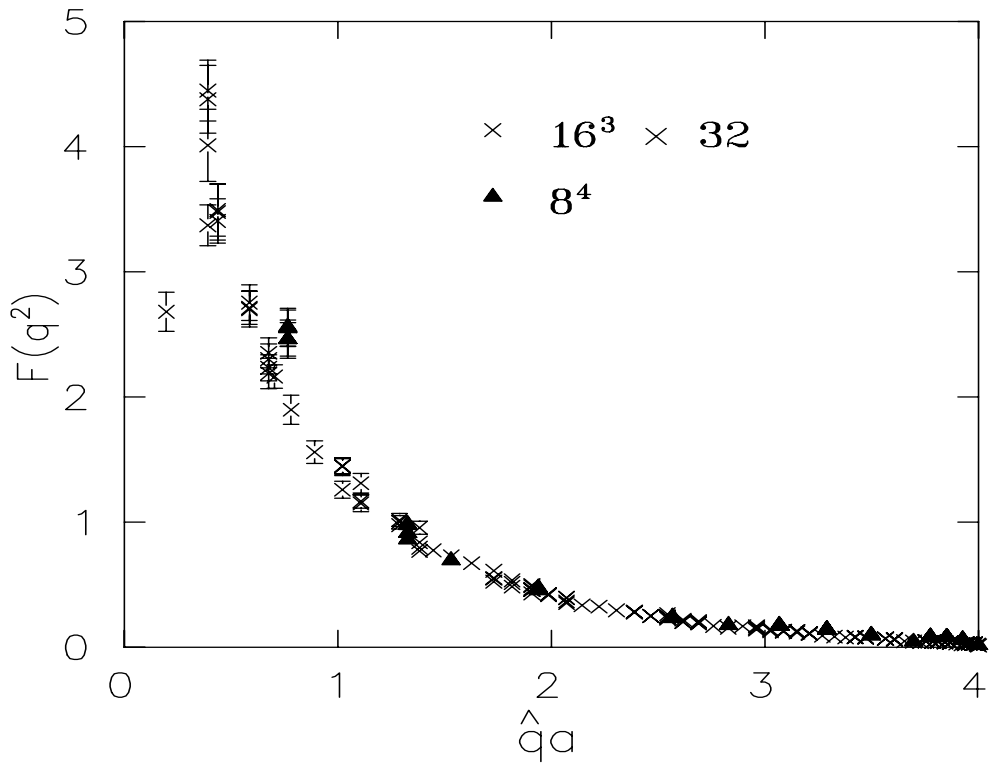}}
\vspace*{0.3cm}
\caption{$F(q^2)$ as a function of $|\hat{q}|$ in lattice units after 
applying the cylindrical cut in momentum.}
\label{fig:xi}
\end{figure}

The difference between Landau and Laplacian gauge 
can also be seen in the deviation of $F(q^2)$ from zero.
Whereas in Landau gauge we find that
\be \hat{q}^\mu \hat{q}^\nu {\cal D}^{aa}_{\mu\nu}\ll 1
\ee
as expected, 
in the Laplacian gauge $F(q^2)$ is not small, and has a maximum at 
low momenta. The behaviour of $F(q^2)$ is shown in Fig.~\ref{fig:xi} for
$8^4$ and $16^3\times 32$ lattices at $\beta=6.0$.  
Since $F(q^2=0)$ can not be obtained by our projection,
we only have one point, at the smallest momentum $2\pi/32$ on the larger
lattice, to ascertain that $F(q^2)$ really has a maximum
and does not keep diverging as $q^2 \rightarrow 0$.
But since the data are systematically higher for the smaller lattice than
for the larger one, it seems unlikely that increasing the lattice size
further would bring the infrared data up and remove the maximum.

It is interesting to examine the volume dependence of the zero-momentum
propagator ${\cal D}(0)\equiv \frac{1}{4}
\sum\nolimits_{\mu\mu}\sum\nolimits_{\alpha}{\cal D}_{\mu\mu}^{\alpha\alpha}(q^2=0)$. 
We note that in order to
determine the transverse part of the propagator at zero-momentum, $D(0)$,
one must subtract from ${\cal D}(q^2)$
$F(q^2)/q^2|_{q^2=0}$, which we can only obtain as 
${\rm lim}_{q^2 \rightarrow 0} F(q^2)/q^2$.

\begin{figure}[htb]
\epsfxsize=13truecm
\epsfysize=10truecm
\mbox{\epsfbox{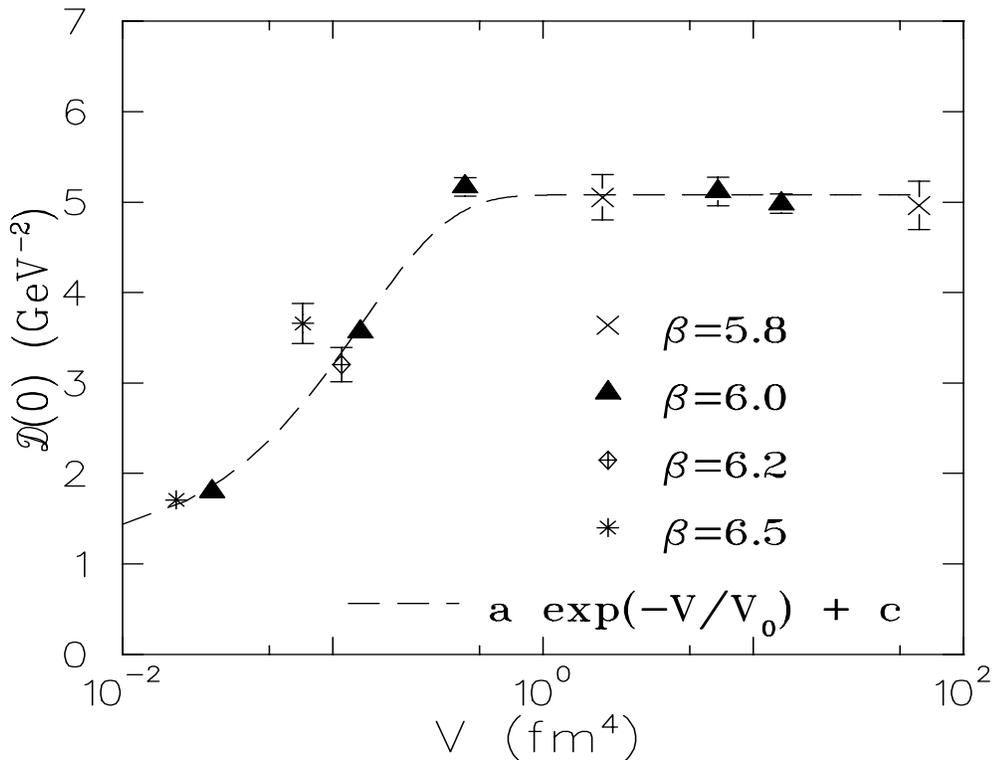}}
\caption{The Renormalized zero-momentum propagator 
${\cal D}(0) \equiv \frac{1}{4}\sum\nolimits_{\mu\mu}\sum\nolimits_{\alpha}
\frac{1}{8}{\cal D}_{\mu\mu}^{\alpha\alpha}(q^2=0)$ 
versus volume in physical units. The dashed line is
a fit to the form $a\exp{(-V/V_0)}+c$.}
\label{fig:D0}
\end{figure} 

>From Fig.~\ref{fig:xi}
it can be seen that 
to extract the limit of $F(q^2)/q^2$ as  
$q^2 \rightarrow  0$ reliably,
one needs more data in the infrared. Therefore, with the volumes at our
disposal we
can only examine the zero-momentum limit ${\cal D}(0)$ of the {\em full} propagator.
In the
Landau gauge, Zwanziger has argued that $D(0) = {\cal D}(0)$ should vanish in the infinite
lattice volume limit \cite{Zwanziger}. 
A recent lattice study in $SU(2)$ at finite temperature~\cite{Cucchieri}
seems indeed to indicate such a behaviour. With the data on our present 
volumes the needed subtraction in the Laplacian gauge
cannot be reliably performed, and thus we cannot extract $D(0)$.
What  we find is that ${\cal D}(0)$,
the zero-momentum propagator,
is finite and volume independent for large enough volumes.
The volume dependence and scaling of the renormalized zero-momentum propagator
 in physical units
 is displayed in Fig.~\ref{fig:D0}
where we collected results from $\beta=5.8,6.0,6.2$ and $6.5$.
To obtain the renormalized propagator we impose 
the renormalization condition  given in 
eq.~(\ref{renormalization point}) where we choose
the renormalization point to be  $\mu=a^{-1}$ for $\beta=6.0$ i.e.
$\mu=1.943$ GeV.
This determines $Z_3(\mu,a_{\beta=6.0}) \approx 2.312$. 
We then use eq.~(\ref{D_R})
to find the ratio of the factors $Z3$ for different $\beta$ values at the
same physical momentum $q$, e.g.
for $\beta=5.8$ at $q=\mu$ we have:
\beq
\frac{Z_3(\mu,a_{\beta=5.8}) }{Z_3(\mu,a_{\beta=6.0})} &=&
 \frac{a_{\beta=5.8}^2
D(qa_{\beta=5.8})}{a_{\beta=6.0}^2 D(qa_{\beta=6.0})} \nonumber \\
 &=&0.97(4) 
\eeq
In this way we obtain the renormalization factors at all  $\beta$ values.
For $\beta=6.2$ and $\beta=6.5$ we find 1.04(4) and 1.07(6) respectively
as compared to the value at $\beta=6.0$.  
We also obtained consistent results by fitting our data in the ultraviolet
regime using the asymptotic one-loop result 
for $D_R\sim Z/q^2 (1/2\ln(q^2/\Lambda^2))^{-d_D}$, with $d_D=13/22$ as in the
Landau gauge since in this regime the results in the Laplacian and Landau
 gauges are the same.
As can be seen 
from Fig.~\ref{fig:D0} the renormalized ${\cal D}(0)$  displays 
reasonable scaling, and appears quite volume-independent 
for volumes larger than $\sim 1/2$ fm$^4$.
We find a value of ${\cal D}(0)=5.020(16)$ GeV$^{-2}$, 
or ${\cal D}(0)^{-1/2}=445(3)$ MeV, 
corresponding to  a length scale of $\sim 0.5$ fm.
Since the zero-momentum propagator
measures the susceptibility of the $A_{\mu}^{a}$ field, the
length associated with it determines the domain over which the 
gluon field remains correlated in the Laplacian gauge. 
If the lattice dimensions become of the order of this
characteristic length, then one expects finite size effects to become appreciable. This is
indeed what is observed, as shown in Fig.\ref{fig:D0}, 
with an approximate volume dependence of 
$exp(-V/V_0)$ with $V$ the lattice volume and $V_0 \sim {\cal D}(0)^2$.

\begin{figure}
\epsfxsize=14truecm
\epsfysize=10truecm
\mbox{\epsfbox{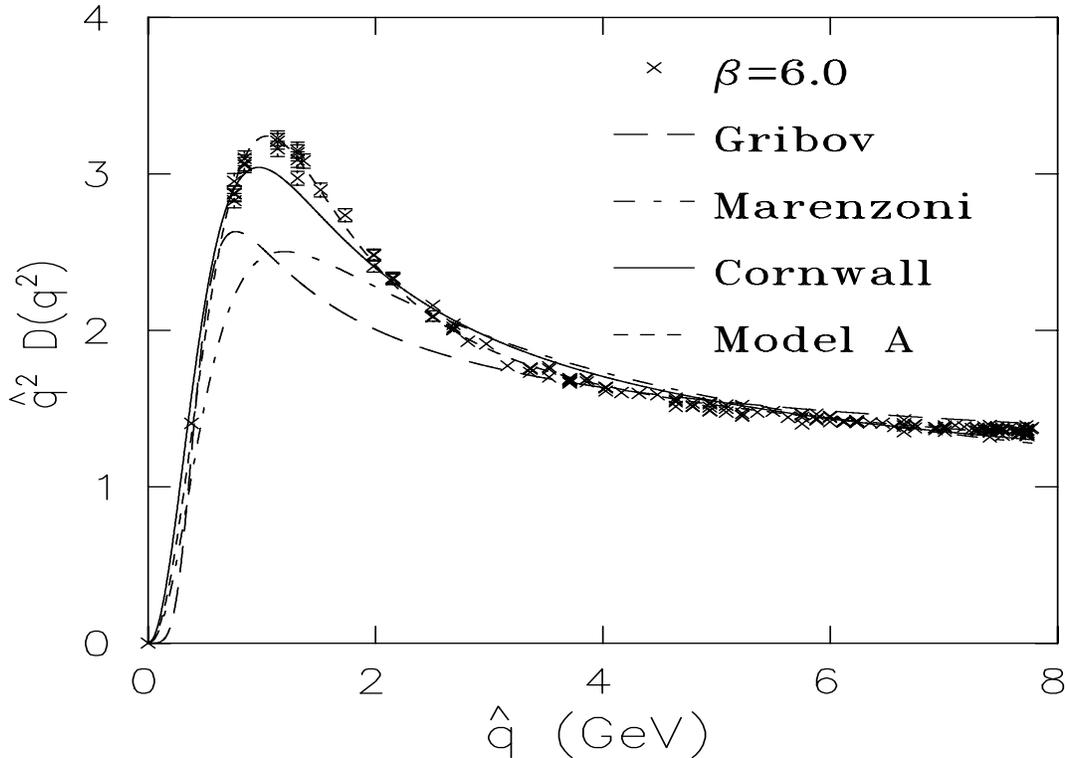}}
\caption{The gluon propagator $D(q^2)$ multiplied 
by $\hat{q}^2$
on the $16^3\times 32$ lattice at $\beta=6.0$. The dashed-dotted line
shows the fit to the model by Marenzoni {\it et al.} eq.(\ref{Marenzoni}),
the solid line to Cornwall's model eq.(\ref{Cornwall}) 
and the dashed line to model A of Ref.~\protect\cite{Leinweber}, eq.(\ref{modelA}).} 
\label{fig:CornI}
\end{figure}

On the lattice, the Lorentz symmetry is only approximately restored.
Lattice artifacts cause some dependence of $D(q)$ on the
orientation of the vector $q$ rather than just on $q^2$. 
To minimize these discretization effects, we
filter our data by making a cylindrical cut in momentum
along a reference direction $\hat{n}=\frac{1}{2}(1,1,1,1)$, 
in the same manner as in Ref.~\cite{Leinweber}.
Namely, we only consider momenta obeying the criterion 
$|\Delta \hat{q}| < 2\pi/L_s$, where $L_s$
is the number of sites in the spatial direction, 
and $\Delta \hat{q}$ is the momentum transverse to $\hat{n}$
($\Delta \hat{q} = \hat{q} - \hat{q}.\hat{n}~\hat{n}$).
Using these filtered data which allow a direct comparison 
with~\cite{Leinweber}, 
we examine the various proposals discussed in the Introduction
for the infrared behaviour of the propagator. 
We find that Gribov type parametrizations \cite{Gribov,Stingl}
as well as infrared enhancement of the type 
$(q^2)^{-2}$~\cite{Mandelstam,BBZ,BP} are excluded~\cite{AFF}.
The ansatz of Marenzoni et al.~\cite{Marenzoni}, 
\be
D(q^2) = \frac{Z}{(q^2)^{1+\alpha} + M^2} \quad,
\label{Marenzoni}
\ee
with a non-perturbative anomalous dimension $\alpha$,
gives a better description of the lattice data
 than the aforementioned parametrizations,
but, as seen in Fig.~\ref{fig:CornI},
 underestimates the peak of the propagator.
On the other hand, 
Cornwall~\cite{Cornwall} allows for a dynamically generated gluon mass 
which vanishes at large momentum in accord with  perturbation theory.
 Using a special set of DSE referred to as a
gauge invariant ``pinch technique'', he obtains the
following solution for the gluon propagator
$$
D(q^2)=Z\> \biggl[ \left(q^2+M^2(q^2)\right){\rm ln}
                             \frac{q^2+4M^2(q^2)}{\Lambda^2}\biggr]^{-1}
$$
with 
\be M(q^2)=M\left\{\frac{{\rm
    ln}\left[(q^2+4M^2)/\Lambda^2\right]}
                {{\rm ln}\left[4M^2/\Lambda^2\right]}\right\}^{-6/11}
\label{Cornwall}
\ee

Cornwall's proposal provides a reasonable fit to the data
over the whole momentum range (with $\chi^2/{\rm n.d.f} = 2.5$).    
The quality
of this fit can be seen in  Fig.~\ref{fig:CornI}.
For comparison we also fitted our data to the form suggested by Leinweber
{\it et al.}~\cite{Leinweber}
 where two terms were used, one to describe the ultraviolet
behaviour of the form  $D_{\rm UV}\sim \frac{1}{q^2+M^2} \>L(q^2,M)$,
and one the
infrared of the form $D_{\rm IR}\sim 1/(q^2+M^2)^{1+\alpha}$. 
The exact form, referred to as model A, as taken from ref.~\cite{Leinweber}, is
\beq
D(q^2)&=&Z
\biggl[\frac{AM^{2\alpha}}{(q^2+M^2)^{1+\alpha}}
                + \frac{1}{q^2+M^2} \>L(q^2,M) \biggl]
 \nonumber \\
L(q^2, M)&= &\biggl\{\frac{1}{2} \rm{ln}\left [
  (q^2+M^2)(q^{-2}+M^{-2})\right] \biggr\}^{-13/22}
\label{modelA}
\eeq
This parametrization, which includes one more parameter than Cornwall's and
is purely phenomenological, does
fit the data best over the whole momentum range 
(with $\chi^2/{\rm n.d.f} = 1.2$).

\begin{figure}
\epsfxsize=13truecm
\epsfysize=10truecm
\mbox{\epsfbox{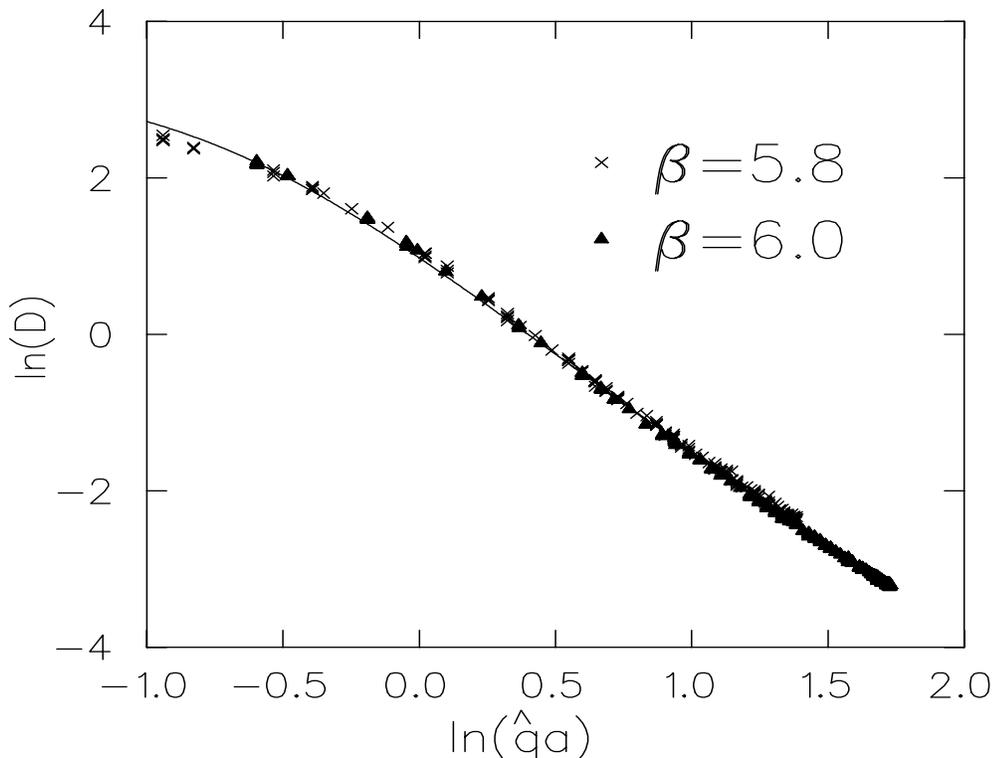}}
\caption{Scaling  of the data at $\beta=5.8$ and $6.0$ on the $16^3\times 32$
lattice. The solid curve is the best fit to both sets of data.} 
\label{fig:scaling}
\end{figure} 

We address the question of scaling by comparing our results at $\beta=5.8$ 
and $\beta=6.0$
on the largest lattice. 
In the scaling regime the renormalized propagator $D_{\rm R}(q;\mu)$ is
independent of the lattice spacing. Therefore, as in ref.~\cite{Leinweber}, 
we can use Eq.~(\ref{D_R}) to obtain the following expression for the ratio of
 unrenormalized  lattice propagators at some physical momentum scale $q$
\be
\frac{D_1(qa_1)}{D_2(qa_2)}=\frac{Z_3(\mu,a_1)D_{\rm R}(q;\mu)/a_1^2}
                                 {Z_3(\mu,a_2)D_{\rm R}(q;\mu)/a_2^2}
                            =\frac{Z_1}{Z_2}\frac{a_2^2}{a_1^2}
\label{renormalization}
\ee
and where the labels $1,2$ refer to the data at  $\beta=6.0$ and $\beta=5.8$
respectively.
 The scaling properties of the lattice gluon propagator can now be investigated
using Eq.~(\ref{renormalization}) by adjusting the ratios $Z_1/Z_2$ and
$a_1/a_2$. 
In Fig.~\ref{fig:scaling} we show
the two sets of data lying on the best scaling curve. The shifts required
along the horizontal and vertical axes determine the
ratios of the wavefunction renormalization constants  and of the 
lattice spacings.
We find 
\be
a_{\beta=6.0}/a_{\beta=5.8} = 0.71 \pm 0.02 \hspace*{2cm}
{\rm and} \hspace*{2cm} Z_{\beta=6.0}/Z_{\beta=5.8} = 1.07 \pm 0.075 \quad.
\ee
with strongly correlated errors.
The ratio of lattice spacings is in  agreement with
the value of $0.72(4)$ obtained from a detailed analysis of the
static potential~\cite{Bali}.
The ratio of the Z-factors is within what is expected from
perturbation theory, and in agreement with the value of $1.04(3)$ of
Ref.~\cite{Leinweber}.
In other words, scaling is very well satisfied for the Laplacian gauge,
and performing the fits  at $\beta=6.0$ gives the behaviour
of the gluon propagator in the physical regime.

\begin{figure}[b]
\epsfxsize=14truecm
\epsfysize=12truecm
\mbox{\epsfbox{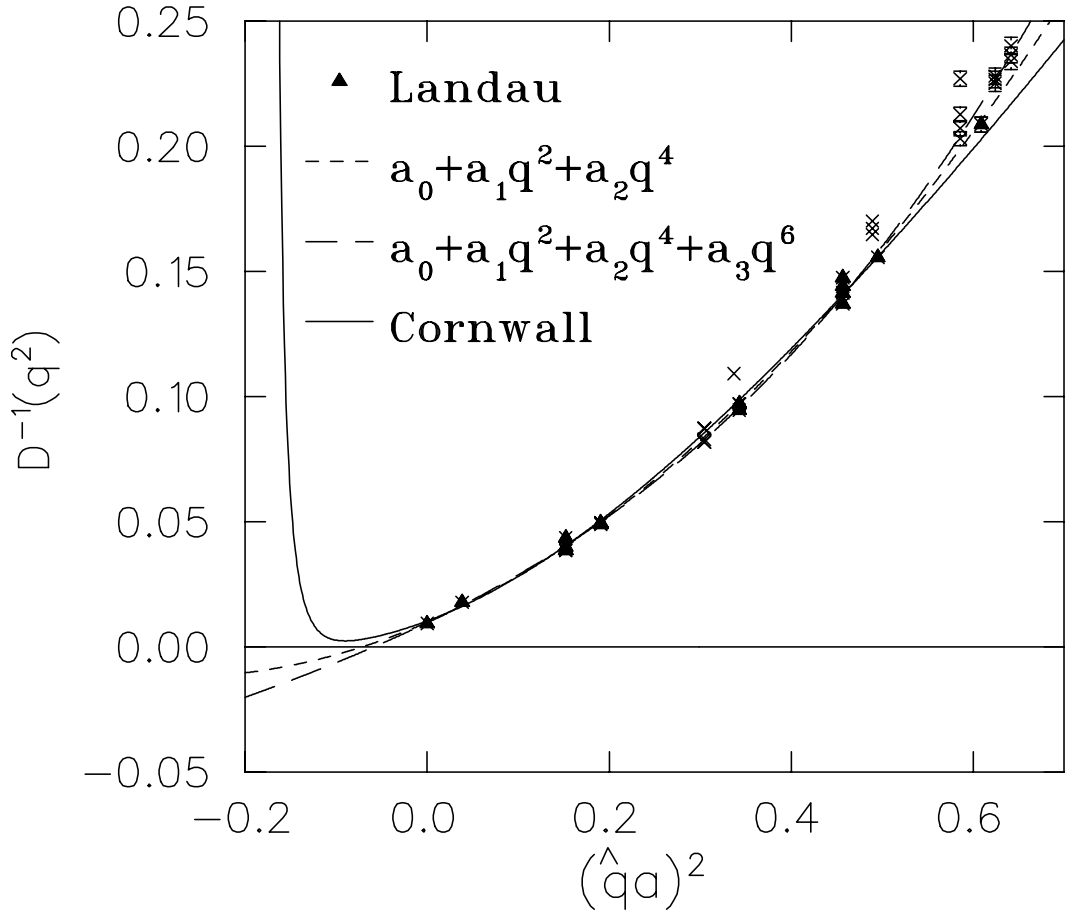}}
\vspace*{0.5cm}
\caption{The inverse gluon propagator at low momentum in Landau gauge,
at $\beta=6.0$ on the $16^3 \times 32$ lattice. 
The filled triangles and  crosses
show the data  which are kept and 
discarded by the cylindrical momentum cut respectively.
Three extrapolations to negative $\hat{q}^2$ are shown: quadratic and cubic 
polynomials in $\hat{q}^2$, and Cornwall's model. Note the instability of the
pole $D^{-1}(q^2)=0$ with respect to the type of extrapolation chosen.}
\label{fig:Dinv_Landau}
\end{figure}

\begin{figure}[b]
\epsfxsize=14truecm
\epsfysize=12truecm
\mbox{\epsfbox{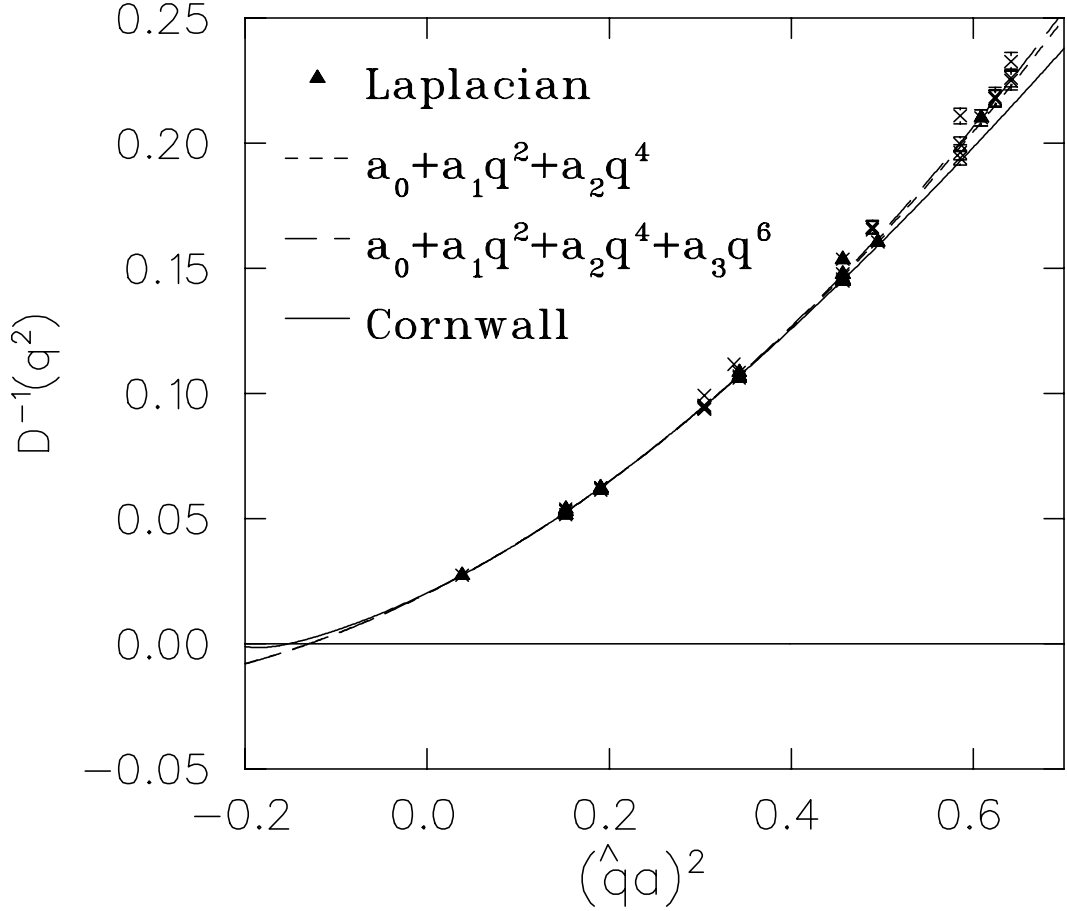}}
\vspace*{0.5cm}
\caption{Same as Fig.\ref{fig:Dinv_Landau}, for the Laplacian gauge.
The reduced vertical scatter of the data at a given momentum indicates
a superior restoration of rotational symmetry. 
The reduced curvature as a function of $\hat{q}^2$ improves the stability of the pole
with respect to the type of extrapolation.}
\label{fig:Dinv_Laplace}
\end{figure}

We focus now on the infrared behaviour of the transverse propagator.
Figs.~\ref{fig:Dinv_Landau} and \ref{fig:Dinv_Laplace} show the
inverse propagator as a function of $\hat{q}^2$ in the two gauges.
Two advantages of the Laplacian gauge become visible.

First, the orientation of the momentum $q$ has less effect   
than in Landau gauge: the data points at a given value of $\hat{q}^2$ show
less scatter, and the cylindrical cut is not as essential as in Landau 
gauge in the infrared region.
At a given lattice spacing,
the Laplacian gauge approximates better 
the Lorentz symmetry of the continuum.
This reduction of lattice artifacts is understandable
since the gauge is fixed by considering the lowest-lying eigenvectors
of the Laplacian, which are the least sensitive to UV-cutoff effects.
In contrast, Landau gauge comes from the iteration of a completely local,
UV-dominated process. Better rotational symmetry allows for better accuracy,
or for the same accuracy on coarser lattices.

Second, the inverse propagator is closer to 
a linear function of $\hat{q}^2$
in Laplacian gauge. If it were the propagator for 
a free boson, it would be described by  a straight line  since
 $1/D(q^2)=Z^{-1}(q^2 + m^2)$.
Having curvature means that one has 
a momentum-dependent effective mass $\Pi(q^2)$.
In particular, the infrared
 mass $\Pi(0)$ and the pole mass $\Pi(q^2)$ such that $(q^2 + \Pi(q^2))=0$
are different.

The latter is of special interest, because of its gauge 
independence at least  to all orders in perturbation theory. 
Finding a pole, i.e. a zero of the inverse propagator, requires the
extrapolation of our data to negative $\hat{q}^2$. 
The less curvature in the
inverse propagator in the infrared,
the more reliable the extrapolation will be.

Three types of extrapolation are displayed in the figures: quadratic and
cubic polynomials in $\hat{q}^2$, and our fit to Cornwall's model.
The location of the pole, and even its existence, are affected
by the choice of extrapolation in the Landau gauge. The coefficients 
$a_1,a_2,a_3$ of the cubic polynomial extrapolation keep increasing,
indicating poor stability. Essentially, no statement about a pole
can be made in that gauge. Differentiating between a cubic fit (which gives
a pole) and a Cornwall-type fit (which doesn't) will require extremely
accurate data on large lattices. 
Extracting the pole is also difficult in the Laplacian gauge but at
least one finds a pole with all the ans\"atze that we tried.
Given the convexity of the data, a lower bound is provided by a linear
fit near $q^2=0$, which defines the (gauge-dependent) infrared mass.
Quadratic and cubic terms in the polynomial extrapolation represent
small corrections of decreasing size. One can thus
make some estimate of the gluon pole mass.
A similar study on a lattice of double size, as was      
considered in Ref.~\cite{Leinweber}, would produce more than four times as 
many points in the same $\hat{q}^2$ interval, and should allow for an accurate
determination of the pole mass.

\begin{figure}[htb]
\epsfxsize=13truecm
\epsfysize=10truecm
\mbox{\epsfbox{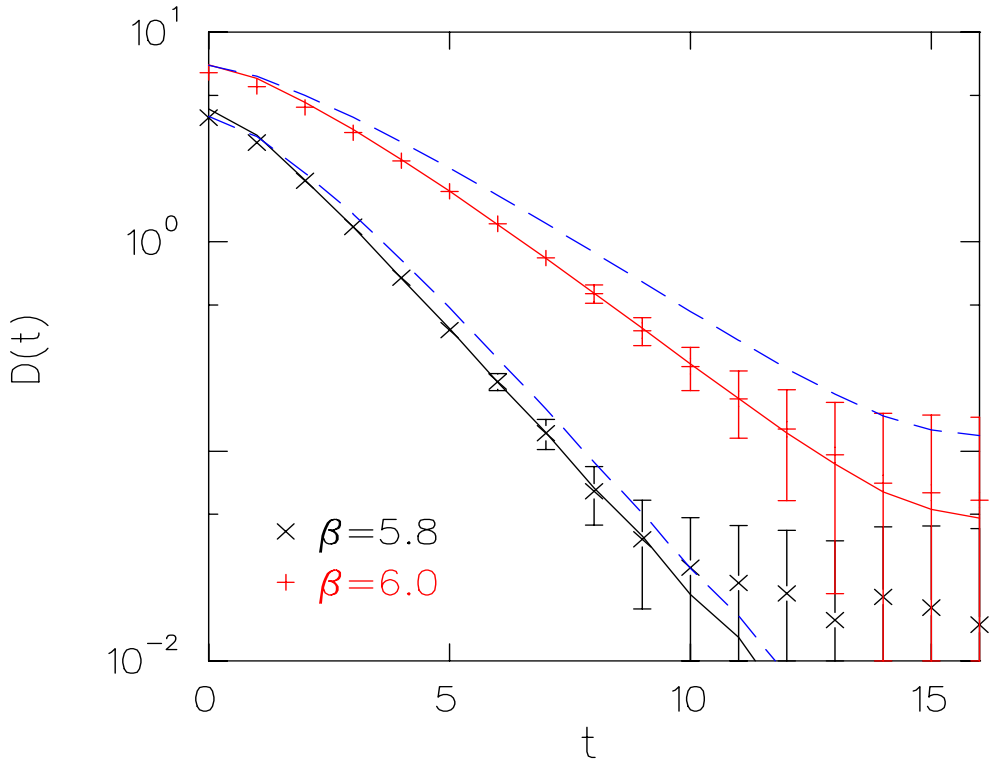}}
\vspace*{0.5cm}
\caption{Time-slice gluon correlator, in Laplacian gauge, at $\beta=5.8$
and $6.0$. The dashed lines show Cornwall's model fitted to $D(\hat{q}^2)$ after
the cylindrical momentum cut; the solid lines are direct fits to the
time-slice correlators, excluding the first few time-slices.}
\label{fig:time-slice}
\end{figure} 

We also measure the correlator of the gluon field averaged over a time-slice.
Namely, we measure
\be
C(t) =\frac{1}{L_s^3} \frac{1}{8} \sum_{a=1}^8 \frac{1}{3} \sum_{\mu=1}^3 
( \sum_x^{L_s^3} A_\mu^a(\vec{x},0) )~( \sum_x^{L_s^3} A_\mu^a(\vec{x},t) )
\label{Ct}
\ee
which is displayed in Fig.\ref{fig:time-slice}.
At large time separations $t$, this correlator should decay exponentially
like $exp(-m_{pole} t)$, giving us another approach 
to extracting the pole mass.
We use this observable to perform a crosscheck on this mass,
and as a further study of the systematic errors in its determination.
This correlator is measured
on the same configurations as $D(q^2)$, so it contains no additional 
information. But the same information is given a different weight, so that
a fit to $C(t)$ will give different results than a fit to $D^{-1}(q^2)$, especially
after the cylindrical momentum cut. Therefore, we fit Cornwall's
model directly to $C(t)$ instead of $D^{-1}(q^2)$. 
Remarkably, the difference is rather small,
which attests again to the soundness of the model. The dashed lines in
Fig.\ref{fig:time-slice} show the original fit of Cornwall's ansatz 
to $D^{-1}(q^2)$, which  already provides a fair description of the data. 
The solid lines represent a direct fit of
the same 3-parameter ansatz to $C(t)$, excluding the first few time-slices
which otherwise completely dominate the fit. 
The fit started from $t=4$ and $t=2$ at $\beta=6.0$
 and 5.8 respectively, which amounts to discarding similar intervals in 
 physical units.
 Given the 3 fitted parameters,
one can then solve $D^{-1}(q^2)=0$ numerically, with $D(q^2)$ as per
eq.(\ref{Cornwall}). The corresponding pole mass varies little from one
fit to the other, and remains roughly constant in physical units at 
$\beta=5.8$ and $6.0$. 
Also, a model-independent extraction of the pole mass, by measuring
the effective mass $m_{eff}(t) = -Ln(C(t+1)/C(t))$, gives a consistent value.
Taking these results
into account, together with the 
quadratic and cubic extrapolations displayed in Fig.\ref{fig:Dinv_Laplace}, 
we estimate the pole mass to lie in the interval $[500,785]$ MeV, where
we used
$a^{-1}(\beta=6.0)=1.943$~GeV to convert to physical units~\cite{Bali} 
with $\sqrt{\sigma}=440$ MeV. 
The lower bound is given by the infrared mass, which corresponds to a linear
extrapolation of $D^{-1}(q^2)$; the upper bound is provided by
the largest value obtained when fitting to our data Cornwall's model.
A reasonable central value is 640 MeV, which corresponds to Cornwall's 
extrapolation in Fig.\ref{fig:Dinv_Laplace}.

We have performed a similar exercise for the Landau gauge.
The fit of Cornwall's model to $D(q^2)$ 
or $C(t)$ is quite satisfactory, 
but the equation $D^{-1}(q^2)=0$  gives a complex pole,
far from the real axis.
Note that Ref.~\cite{karsch} also finds oscillatory behaviour for the
time-slice correlator in $3d$ $SU(2)$ 
theory fixed to Landau gauge, reflecting
a complex pole.
This disagreement with the Laplacian gauge is puzzling,
since one expects the pole to be gauge invariant.
Possible causes include the inadequacy of the Landau gauge fixing procedure
on the lattice, or finite-size effects. Larger volume studies, currently
under way, should elucidate this issue.

\section{Conclusions}

 We have evaluated the gluon propagator using the Laplacian gauge
 which avoids lattice Gribov copies. We extracted the transverse part
of the gluon propagator and verified its scaling in this
gauge. Examining the scaling and volume dependence of the
zero-momentum propagator ${\cal D}(0)$, we reached
the conclusion that it is a constant beyond a lattice size of $\sim 0.8$~fm. 
This size
is consistent with the characteristic length scale determined 
from ${\cal D}(0)$ itself
as the range beyond which the gluon field decorrelates in this gauge.

Among the various proposals for the transverse propagator which are physically founded, 
Cornwall's model~\cite{Cornwall} provides a reasonable 
fit to the lattice results over the whole momentum range.
We find it satisfying that the lattice data seem to favour a model with a 
dynamically generated mass.

By looking at the inverse propagator $D^{-1}(q^2)$ at small momenta, 
we see that the Laplacian gauge is superior to the Landau gauge in its
restoration of Lorentz symmetry on the lattice. Furthermore, it turns out that
the inverse propagator is almost linear in $\hat{q}^2$ in the Laplacian gauge.
This allows for a more reliable extrapolation to $\hat{q}^2 < 0$, as compared to
the Landau gauge. We test a variety of extrapolation ans\"atze. 
They consistently yield a  pole mass at $\sim 640 \pm 140$ MeV.

\vspace*{1cm}
\noindent
\underline{Acknowledgements:}
The $16^3\times 32$ lattice configurations were obtained from the
Gauge Connection archive~\cite{connection}.
We thank A. Cucchieri, S. D\"urr, M. Pepe and O. Philipsen for helpful discussions.

\end{document}